\documentclass[a4paper,11pt]{article}  
\pdfoutput=1 
\usepackage{jcapalt} 
\usepackage[T1]{fontenc} 

\newcommand{\be}{\begin{equation}}
\newcommand{\ee}{\end{equation}}
\newcommand{\simless}{\lower.5ex\hbox{$\; \buildrel < \over \sim\;$}}
\newcommand{\simgreat}{\lower.5ex\hbox{$\; \buildrel > \over \sim\;$}}

\newcommand{\mplanck}{ M_{\rm pl}}

\newcommand{\tempeq}{ T_{\rm eq} }
\newcommand{\timeeq}{ t_{\rm eq} }
\newcommand{\rhoeq}{\rho_{\rm eq}} 
\newcommand{\mpro}{ m_{\rm p} }

\newcommand{\tvir}{t_{\rm vir}} 
\newcommand{\fvir}{f_{\rm vir}} 
\newcommand{\rhovac}{\rho_{\scriptscriptstyle\Lambda} } 
\newcommand{\timevac}{t_{\scriptscriptstyle\Lambda} } 
\newcommand{\timebbn}{t_{\scriptscriptstyle{BBN}} }

\title{\boldmath Constraints on Vacuum Energy from 
Structure Formation and Nucleosynthesis}

\author[1,2]{Fred C. Adams,} 
\author[3]{Stephon Alexander,}  
\author[1]{Evan Grohs,}
\author[4]{ $\qquad$ $\qquad$ and Laura Mersini-Houghton}  

\affiliation[1]{Physics Department, University of Michigan, Ann Arbor, MI 48109} 
\affiliation[2]{Astronomy Department, University of Michigan, Ann Arbor, MI 48109} 
\affiliation[3]{Physics Department, Brown University, Providence, RI 02912} 
\affiliation[4]{Physics Department, University of North Carolina, Chapel Hill, NC 27599}  

\emailAdd{fca@umich.edu} 

\abstract{This paper derives an upper limit on the density $\rhovac$ 
of dark energy based on the requirement that cosmological structure
forms before being frozen out by the eventual acceleration of the
universe. By allowing for variations in both the cosmological
parameters and the strength of gravity, the resulting constraint is a
generalization of previous limits.  The specific parameters under
consideration include the amplitude $Q$ of the primordial density
fluctuations, the Planck mass $\mplanck$, the baryon-to-photon ratio
$\eta$, and the density ratio $\Omega_M/\Omega_b$.  In addition to
structure formation, we use considerations from stellar structure and
Big Bang Nucleosynthesis (BBN) to constrain these quantities. The
resulting upper limit on the dimensionless density of dark energy
becomes $\rhovac/\mplanck^4<10^{-90}$, which is $\sim30$ orders of
magnitude larger than the value in our universe
$\rhovac/\mplanck^4\sim10^{-120}$.  This new limit is much less
restrictive than previous constraints because additional parameters
are allowed to vary. With these generalizations, a much wider range of
universes can develop cosmic structure and support observers. To
constrain the constituent parameters, new BBN calculations are carried
out in the regime where $\eta$ and $G=\mplanck^{-2}$ are much larger
than in our universe. If the BBN epoch were to process all of the
protons into heavier elements, no hydrogen would be left behind to
make water, and the universe would not be viable. However, our results
show that some hydrogen is always left over, even under conditions of
extremely large $\eta$ and $G$, so that a wide range of alternate
universes are potentially habitable. }

\begin{document}
\maketitle
\flushbottom

\section{Introduction} 
\label{sec:intro} 

Our current understanding of the universe falls in a curious regime.
On one hand, we can explain the basic properties of the universe with
relatively simple equations and relatively few basic cosmological
parameters. Observations of the Cosmic Microwave Background Radiation
\cite{wmap} and the expansion rate \cite{riess,perlmutter} provide us
with a description of the universe with a precision approaching a few
percent \cite{riess2}.  On the other hand, one of the required
parameters in this paradigm is the energy density of the vacuum,
denoted here as $\rhovac$ and often called the dark energy. Existing
observations indicate that the density of the dark energy is nearly
constant with time, so that it acts much like a cosmological constant.
In addition, the value of the vacuum energy density is comparable to
the closure density of the universe at the present cosmological epoch,
so that the vacuum energy density is much smaller than the benchmark
value given by the Planck mass, i.e.,
\be
{\rhovac\over\mplanck^4}\sim10^{-120}\ll1\,.
\ee 
This expression, and the rest of the paper, is written in units where
$\hbar=c=k_B=1$.  One would like an explanation for this extreme
ordering of energy scales, but no general consensus currently exists
\cite{weinberg89}. 

In the absence of a definitive prediction for the energy density of
the vacuum, many researchers have argued that the value of $\rhovac$
is constrained by anthropic considerations \cite{carr,carter,bartip}.
In this context, the value of $\rhovac$ and other fundamental
constants must have values that allow for the formation of structure
in the universe and the possibility of the existence of observers.  
An upper bound on the energy density of the vacuum can be derived from
the requirement that galaxy formation occurs before the the universe
becomes dominated by the cosmological constant. The first treatment of
this problem \cite{weinberg87} found that the upper bound on the
vacuum energy density must be at least as large as
$500\rho_0>\rhovac$, where $\rho_0$ is the current density of the
universe. This value was obtained under that assumption that quasars
must form by redshift $z=4.5$.  Weaker bounds --- corresponding to
larger values of the upper bound --- can be derived by relaxing the
assumption that structures must form by the current epoch or that the
perturbations must begin with the small amplitudes realized in our
universe \cite{aguirre,martel,mersini}. These generalizations allow
the upper limit to be much larger, i.e., $\rhovac<10^9\rho_0$. 
Given that the vacuum energy density could be a billion times larger
than its observed value, anthropic arguments are not overly
constraining. 

The goal of this paper is to reexamine the upper limit on the density
of dark energy for a more general class of universes by allowing
additional parameters to vary. Here we consider variations in the
amplitude $Q$ of the primordial density fluctuations, the
baryon-to-photon ratio $\eta$, the total matter density $\Omega_M$,
the baryon density $\Omega_b$, and the strength of gravity (given by
the Planck mass $\mplanck$).  Because different values for the
gravitational constant (equivalently, $\mplanck$) affect stellar
structure, we require that stars are operational, which enforces an
upper limit on the strength of gravity (a lower limit on $\mplanck$).
The values of $\eta$ and $\mplanck$ affect the yields from Big Bang
Nucleosynthesis (BBN). Here we carry out new BBN calculations in the
regime where the values of both $\eta$ and $\mplanck$ are much larger
than in our universe, and use the results to enforce the requirement
that enough hydrogen remains to provide water. These calculations
explore a new regime of parameter space and illustrate the difficulty
of rendering the universe lifeless due to the BBN epoch.

A secondary goal of this paper is to consider the overarching issue of
the possible fine-tuning of the universe
\cite{barnes2012,dirac,tegrees,tegmark}.  It is often claimed that even
small variations in the fundamental constants of physics and/or the
parameters of cosmology would make it impossible for the universe to
develop complex structures and hence observers
\cite{bartip,carr,hogan,reessix}.  Here we explicitly consider
the possible range of values for $\rhovac$. In addition, in order to
evaluate our constraint on the energy density of the vacuum, we must
consider the range of possible variations for the fluctuation
amplitude $Q$, the baryon to photon ratio $\eta$, the Planck mass
$\mplanck$, and the ratio $\Omega_M/\Omega_b$. We find that all of
these quantities, as well as the vacuum energy density $\rhovac$
itself, can vary over wide ranges without rendering the universe
inhospitable.

This paper is organized as follows. We first provide a brief review of
the basic elements of structure formation and then derive the
constraint on $\rhovac$ resulting from the requirement that structure
form before the universe becomes vacuum dominated (Section
\ref{sec:derive}).  The resulting constraint depends on the values of
the parameters $(Q,\eta,\mplanck,\Omega_M/\Omega_b)$; the allowed
range of these parameters are discussed and constrained in Section
\ref{sec:evaluate}.  The paper concludes in Section \ref{sec:conclude}
with a summary of the results and a discussion of their implications.

\section{Constraint on the Energy Density of the Vacuum} 
\label{sec:derive} 

\subsection{Definitions} 
\label{sec:define} 

The equation of motion for the scale factor $a(t)$ of the universe 
has the form 
\be
\left( {{\dot a} \over a} \right)^2 = {8 \pi G \over 3} 
\left[ \rhovac + \rho_M + \rho_R \right] \,, 
\ee
where we have assumed that the universe is spatially flat, which would
be the case if inflation occurs under standard conditions. The densities 
on the right-hand-side of the equation correspond to the vacuum energy 
($\rhovac$), matter ($\rho_M$), and radiation ($\rho_R$). For the sake  
of definiteness, we assume that the universe experiences both 
radiation dominated and matter dominated epochs.  Suppose that the
universe has total energy density $\rhoeq$ at the epoch of equality.
If we use this value as a reference density scale, then we can 
define a corresponding reference time scale 
\be
\timeeq \equiv \left({8\pi G \rhoeq \over 3}\right)^{-1/2}\,,
\label{tbench} 
\ee
and the equation of motion reduces to the form  
\be
\left( {{\dot a} \over a} \right)^2 = \Omega_V + 
\Omega_M a^{-3} + \Omega_R a^{-4} \,,
\ee
where $\Omega_V=\rhovac/\rhoeq$, $\Omega_M=\rho_M/\rhoeq$, and
$\Omega_R=\rho_R/\rhoeq$.  The time variable in the reduced equation
of motion is related to the physical time $t_{\rm phys}$ (coordinate 
time) according to $t=t_{\rm phys}/\timeeq$. Flatness of the universe
also implies the constraint 
\be
\Omega_V + \Omega_M + \Omega_R = 1 \,.
\ee
If we also let the scale factor $a=1$ at the epoch of equality, then
$\Omega_M = \Omega_R \equiv \Omega$, and we redefine $\lambda$ =
$\Omega_V=1-2\Omega$. The equation of motion now becomes
\be
\left( {{\dot a} \over a} \right)^2 = \lambda + 
\Omega \left[ a^{-3} + a^{-4} \right] \,,
\ee
where $\Omega \approx 1/2$ and $\lambda\ll1$. 
For future reference, the temperature of the universe at the 
epoch of equality can be written in the form \cite{tegrees,tegmark}
\be
\tempeq = \eta \, \mpro {\Omega_M \over \Omega_b} \,,
\label{tequal} 
\ee
where the parameter $\eta$ is the baryon to photon ratio and $\mpro$
is the proton mass. Notice also that we are ignoring the neutrino
contribution to the energy density of radiation.

\subsection{Review of Perturbation Growth} 
\label{sec:growth} 

The usual equation of motion \cite{kolbturn} for the growth of density
fluctuations $\delta_k=\rho_k/\rho_0$ in an expanding universe has the
form 
\be
{\ddot \delta_k} + 2 {{\dot a} \over a} {\dot \delta_k} 
+ \left[ {v_s^2 k^2 \over a^2} - 4 \pi G \rho_M \right] \delta_k 
= 0 \,.
\ee
This equation is written in physical units, where $\rho_k$ is the
density perturbation, $k$ is the wavenumber of the perturbation, and
$\rho_M$ is the density of the matter. Here we consider only long 
wavelength modes $k\ll k_J$, where $k_J$ is the Jeans wavenumber, so
that we can drop the subscript and ignore the pressure term. 
Converting to the dimensionless units of the previous section, using
the reference time scale $\timeeq$ and density scale $\rhoeq$, the
equation of motion for $\delta$ becomes 
\be
{\ddot \delta} + 2 {{\dot a} \over a} {\dot \delta} 
- {3 \over 2} \Omega a^{-3} \,\delta = 0 \,.
\label{dequation} 
\ee
The remaining equation of motion for the Hubble expansion has 
the form 
\be
\left( {{\dot a} \over a} \right)^2 = \lambda  + 
\Omega \left[ a^{-3} + a^{-4} \right] \,, 
\label{aequation} 
\ee
where $\lambda\equiv1-2\Omega$. By definition, growth starts at
$t=1$, when the initial conditions are 
\be
a = 1, \qquad \delta = 1, \qquad {\rm and} \qquad 
{\dot \delta} = 0\,.
\label{incon} 
\ee
Note that since the equation of motion for the density perturbation is
linear, we can assume any value for the starting state and re-scale it
later. 

\begin{figure}[tbp]
\centering 
\includegraphics[width=.90\textwidth,trim=0 150 0 150,clip]{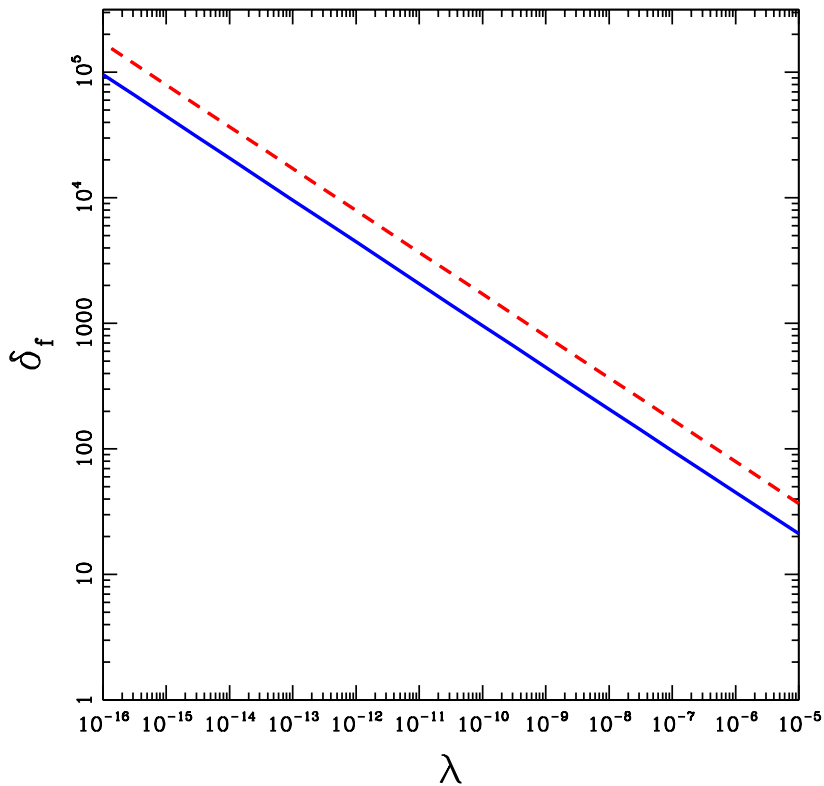}
\caption{Growth factor as a function of the vacuum energy density, 
expressed here as the fraction of the total energy density at the 
time of equality (between radiation and matter). The solid blue curve 
shows the result from numerical integration, whereas the dashed red curve 
shows the approximate result $\delta_{\rm f}$ = $(2\lambda)^{-1/3}$
(see text). }
\label{fig:growth} 
\end{figure}

With the above formulation, we have a one parameter family of models
for structure formation. Moreover, as long as $\lambda \ne 0$, the
perturbation $\delta \to \delta_{\rm f}$ = {\sl constant} in the limit
$t\to\infty$. For the initial condition $\delta(t=1)=1$, the final
value $\delta_{\rm f}$ thus represents the total amount of growth
available for a given value of the vacuum contribution. This growth
factor is shown in Figure \ref{fig:growth}, plotted here as a function
of $\lambda$. To leading order, we expect the amplitude of a density 
perturbation to grow linearly with the scale factor $a(t)$ during 
the matter dominated era and then to saturate (freeze out) when the 
universe becomes vacuum dominated. The total growth factor should 
be approximately $\delta_{\rm est} \sim (2\lambda)^{-1/3}$. This factor 
is shown as the dashed curve in Figure \ref{fig:growth}. Note that 
the approximate expression (dashed curve) closely follows the more 
exact value (solid curve) calculated by numerically evaluating 
equations (\ref{aequation}) and (\ref{dequation}) subject to the 
initial conditions of (\ref{incon}). 

Notice also that the linear treatment is approximate. We can correct 
for this deficiency by writing the growth factor in the form
\be 
\delta_{\rm est} = A \lambda^{-1/3} \,,
\label{growth} 
\ee
where $A$ is a dimensionless parameter of order unity. In the original 
derivation of this constraint \cite{weinberg87}, this parameter is given 
by $A=(500/729)^{1/3}\approx0.882$. Although we include the factor $A$ 
for completeness here, the exact value will not matter: as shown below, 
the bound changes by 30 orders of magnitude. 

\subsection{Constraint on Vacuum Energy}
\label{sec:convac} 

We can now write down the constraint due to the requirement that
cosmological structures can form. The initial size of perturbations is
set by the parameter $Q$, and the growth factor is given by equation
(\ref{growth}). In order for the perturbations to become nonlinear
before being frozen out due to the acceleration of the expansion, we
require 
\be
\lambda < A^3 Q^3 \,. 
\label{conzero} 
\ee
In our universe the parameter $Q \sim 10^{-5}$ \cite{cobe,wmap}.  
In any universe, we expect $Q\ll1$ in order for habitable galactic
structures to form \cite{tegrees,tegmark,reessix,coppess}, so that the
above constraint also implies $\lambda\ll1$.  As a result, for the
time scales and temperature scales evaluated at the epoch of equality,
any corrections due to the nonzero value of $\lambda$ must be small.
Notice also that this constraint is essentially the same as that given
by equation (29) of \cite{tegrees}, or by equation (4) of
\cite{martel}.

Recall that $\lambda$ is the fractional contribution of the 
vacuum to the energy density at the epoch of equality. It is useful 
to write the constraint (\ref{conzero}) in terms of the (constant) 
energy density $\rhovac$ of the vacuum, i.e., 
\be
{\rhovac \over \mplanck^4} \le A^3 Q^3 {\rhoeq \over \mplanck^4} 
= (2 A^3 a_R) Q^3 \eta^4 \left( {\mpro \over \mplanck} \right)^4 
\left( {\Omega_M \over \Omega_b} \right)^4 \,, 
\label{vacbound}
\ee
where we have scaled $\rhovac$ by its ``natural'' value implied by the
Planck mass. To obtain the final equality, we have used the expression
$\rhoeq = 2 a_R \tempeq^4$, where $a_R=\pi^2/15$ is the radiation
density constant, along with the expression (\ref{tequal}) for the
crossover temperature $\tempeq$.  

In our universe, $\mpro/\mplanck\sim10^{-19}$, $\eta\sim10^{-9}$,
$\Omega_M\sim6\Omega_b$, and $Q\sim10^{-5}$, so the dimensionless
vacuum energy $\rhovac/\mplanck^4\sim10^{-124}$. In order to place 
an upper bound on this quantity, we need to separately constrain 
the possible ranges for the parameters $Q$, $\eta$, $\mplanck$, 
and $\Omega_M/\Omega_b$. 

\section{Limits on the Input Parameters}
\label{sec:evaluate} 

As outlined above, if we use the values of $Q$, $\eta$, and $\mplanck$
found in our universe, then the bound on the vacuum energy density
from equation (\ref{vacbound}) leads to a value of $\rhovac$ roughly
comparable to that inferred by observational data. Instead of using
the observed values for $(Q,\eta,\mplanck)$, however, we instead need
to find upper bounds on these quantities. These bound are discussed
below.

\subsection{Density Fluctuation Amplitude} 
\label{sec:qconstaint} 

Constraints on the fluctuation amplitude $Q$ have been considered by 
several previous authors \cite{tegrees,coppess}. If the fluctuation
amplitude is larger, then galaxies form earlier in cosmological
history, when the background density is larger. This ordering of time
scales results in galaxies that are denser. If galaxies are too
dense, then planets in habitable orbits can be disrupted through
scattering interaction with passing stars \cite{tegrees}. Since
galaxies have a wide range of densities, however, this effect does not
render the entire galaxy uninhabitable. Instead, it limits the
fraction of stars that could in principle harbor habitable planets
\cite{coppess}. If we take an optimistic view, the fluctuation
amplitude $Q$ could be as large as $Q\sim10^{-2}$, which implies that
about half of the stars in a galaxy the size of our Milky Way would
remain habitable. This value of $Q$, in turn, allows the bound in
equation (\ref{vacbound}) to be larger (than for the parameters in our
universe) by a factor of $10^9$.

\subsection{Baryon to Photon Ratio} 
\label{sec:etaconstaint} 

Bang Nucleosynthesis (BBN) provides a constraint on the value of the
baryon to photon ratio $\eta$. In our universe, this ratio must be
$\eta\sim10^{-9}$ in order for the early universe to produce (roughly)
the observed abundances of the light elements (deuterium, lithium, and
helium). The abundances of these nuclear species could be different in
other universe, varying by large factors, with no detrimental effects.
A universe could end up sterile, however, if the BBN epoch processes
all of the hydrogen into helium or other heavier nuclei. Such a
universe, with little or no hydrogen, would not have the basic raw
materials to make water.

Calculations of the BBN epoch show that the baryon-to-photon ratio
$\eta$ can be increased by many orders of magnitude and still allow
substantial hydrogen to remain unprocessed. Here we present results
computed using the {\sl BURST} code \cite{burst} for BBN, which is
updated from the standard version of the BBN code
\cite{wagoner,kawanocode}. To start, we fix all of the parameters at
their standard values but allow variations in the value of the
baryon-to-photon ratio $\eta$. Note that the gravitational constant
$G$ is varied in the following section, and then both $\eta$ and $G$
are allowed to vary at the same time.

\begin{figure}
\begin{center}
\includegraphics[scale=0.70]{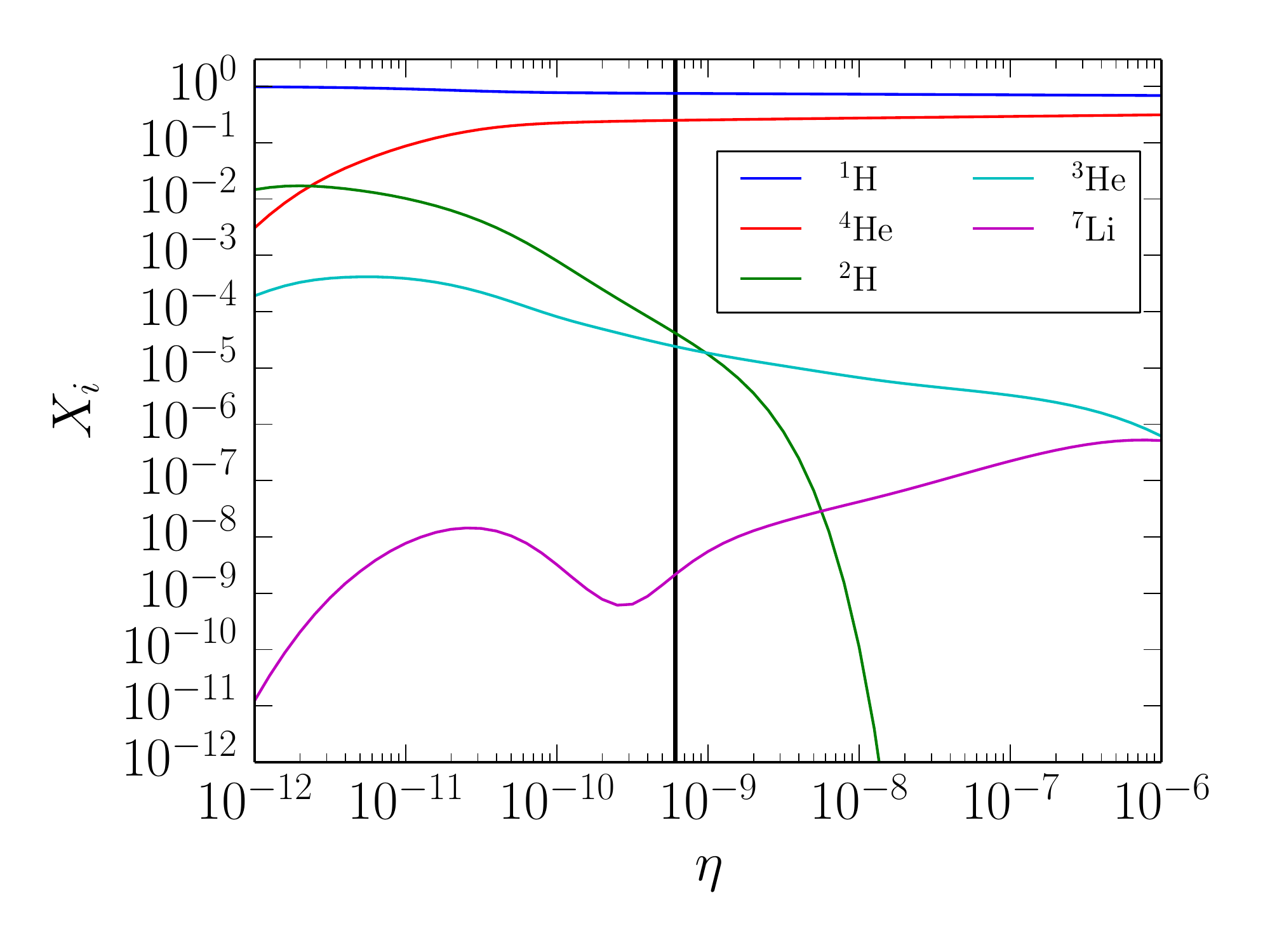}
\end{center} 
\vskip-1.0truecm 
\caption{BBN yields as a function of the baryon to photon ratio $\eta$.
The curves show the resulting mass fraction at the end of the BBN epoch 
for helium-4 (red), as well as the corresponding mass fractions $X_i$ for 
deuterium (green), helium-3 (cyan), and lithium-7 (purple). The blue 
curve shows the mass fraction of free protons. The vertical black line 
marks the estimated value of $\eta=6\times10^{-10}$ for our universe. }
\label{fig:density} 
\end{figure}

Figure \ref{fig:density} shows the resulting BBN yields for a range of
$\eta$ values.  For small values of $\eta$, much smaller than the
values for our universe, the mass fraction of helium-4 is small ---
even smaller than that of deuterium.  As $\eta$ increases, the nuclear
reaction rates increase, and the mass fraction of helium-4 increases.
The abundances of deuterium and helium-3 decrease with increasing
$\eta$, as they are burned into helium-4.

In the limit of large $\eta$, the mass fraction of helium-4 reaches a
limiting value of $Y_4\sim0.3$. This limit corresponds to the regime
where essentially all of the neutrons are burned into helium-4. Notice
also that the abundances of the other nuclear species decrease with
increasing $\eta$. With all of the neutrons incorporated into
helium-4, none are left for constructing the remaining light
elements. For example, with $\eta=10^{-6}$, after helium-4 the next
most abundant nuclear species is helium-3, with a mass fraction of
only $3\times10^{-7}$. As a result, a large fraction of protons always
remains to potentially be incorporated into water. We can thus
increase the value of $\eta$ by a factor of at least $10^3$ and allow
the universe to remain viable. This increase, in turn, allows the
limit of equation (\ref{vacbound}) to increase by a factor of
$10^{12}$.

\subsection{Gravitational Constant (Planck Mass)} 
\label{sec:gconstaint} 

Stellar considerations show that the the gravitational constant can be
larger than the value in our universe, but ``only'' by a factor of
$\sim2\times10^5$ \cite{adams,adamsnew}.  This limit arises from the
requirement that working stars exist, more specifically that stable
nuclear burning states exist, the stellar mass is larger than the
minimum value enforced by degeneracy pressure, and that the lower mass
limit for stars does not exceed the upper mass limit.  The constraint
of equation (\ref{vacbound}) is proportional to $G^2\sim\mplanck^{-4}$, 
so this limit can be increased by a factor of $\sim4\times10^{10}$ due
to the allowed range of the gravitational constant.

\begin{figure}
\begin{center}
\includegraphics[scale=0.70]{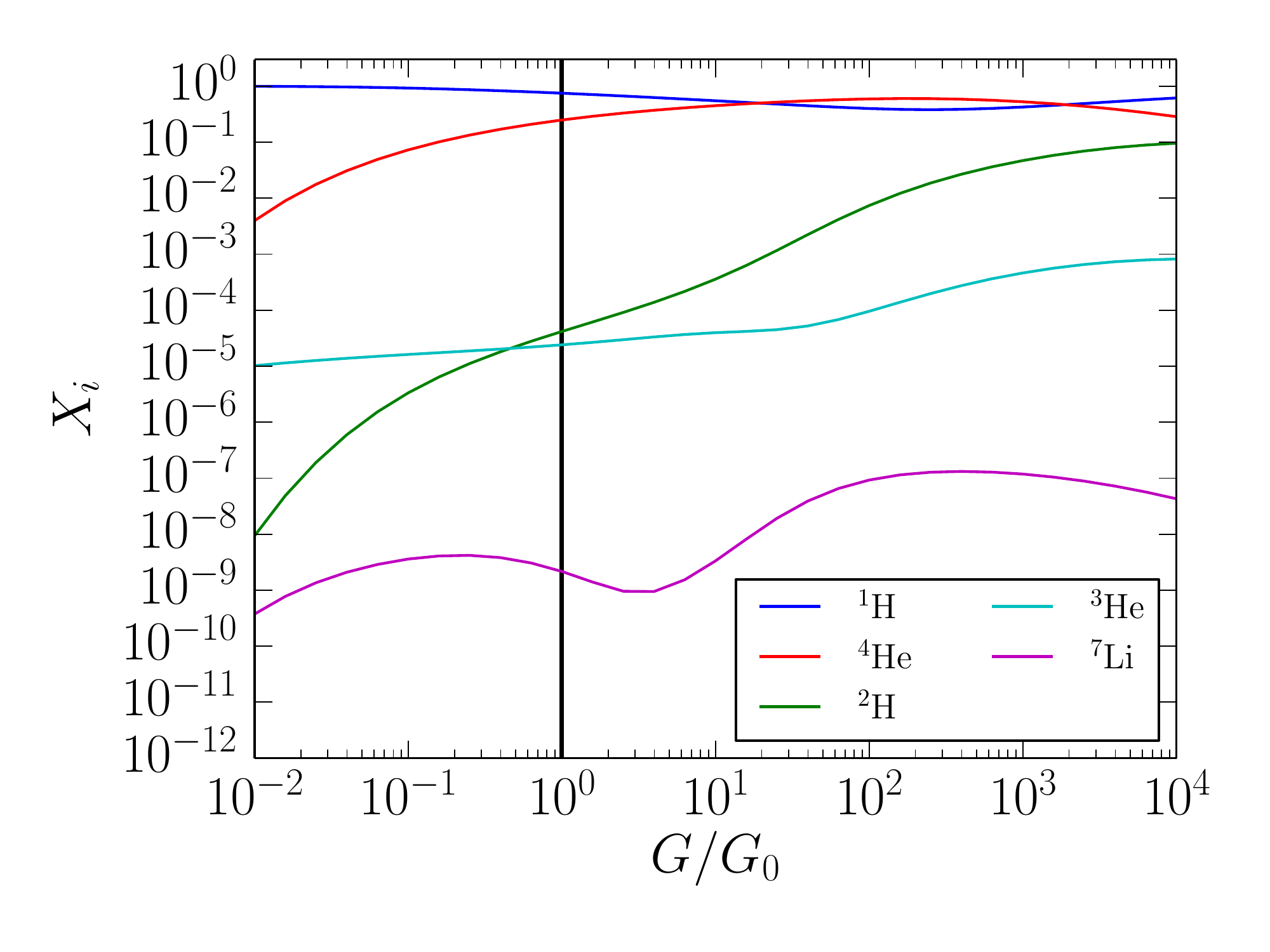}
\end{center}
\vskip-1.0truecm 
\caption{BBN yields as a function of the gravitational constant,
$G/G_0$, scaled to the value in our universe.  The curves show the  
resulting mass fraction at the end of the BBN epoch for helium-4
(red), as well as the corresponding mass fractions $X_i$ for  
deuterium (green), helium-3 (cyan), and lithium-7 (purple). The blue 
curve shows the mass fraction of free protons. The vertical black line 
marks the value of $G$ found in our universe. } 
\label{fig:gravity} 
\end{figure}

Possible variations in the gravitational constant can also change the
predictions of BBN due to corresponding changes in the expansion rate
of the universe, where $H={\dot a}/a\propto{G^{1/2}}$. Figure
\ref{fig:gravity} shows the yields from BBN, where the gravitational
constant is varied over six orders of magnitude, from 100 times
smaller than the value in our universe to $10^4$ times larger. For
most of the range shown, the abundances of all of the light elements
increase with $G$.  For sufficiently large values of $G$, however, the
expansion rate is so fast that not all of the neutrons can be made
into helium-4. As a result, the mass fraction of helium-4 has a
maximum value of $Y_4\sim0.53$, which occurs at $G/G_0\sim100$. Even
with this maximum mass fraction of helium-4, the universe retains
about half of its protons to make water. As result, BBN does not
greatly constrain the habitability of universes in this context.

\begin{figure}
\begin{center}
\includegraphics[scale=0.70]{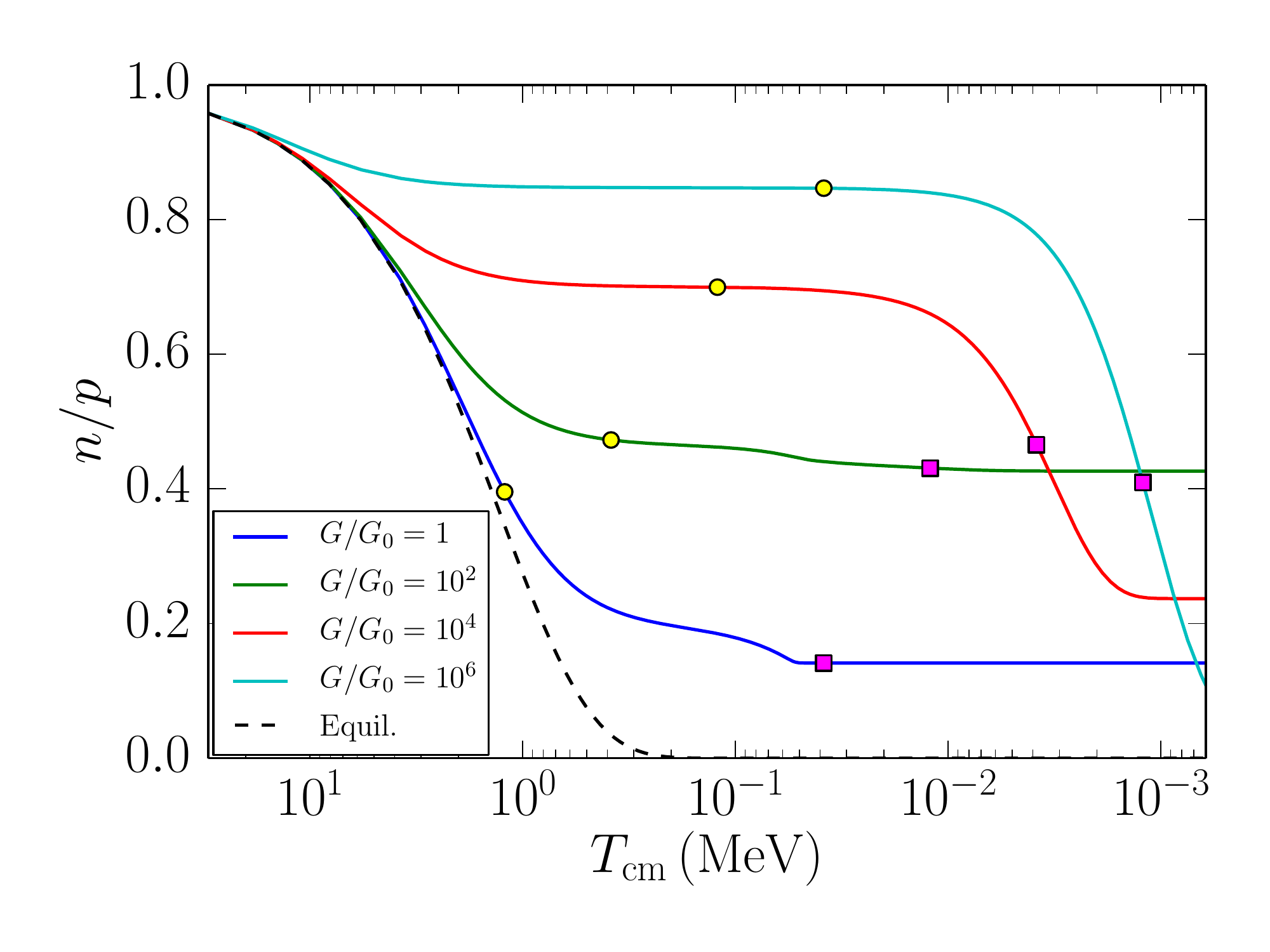} 
\end{center}
\vskip-1.0truecm 
\caption{Neutron to proton ratio during the BBN epoch. The ratio $n/p$ 
is plotted versus the temperature parameter $T_{\rm cm}$ for different 
choices of the gravitational constant, as labeled (see Ref.
\cite{grohsetal} for a precise definition of $T_{\rm cm}$). The black
dashed curve shows the $n/p$ ratio expected in Nuclear Statistical
Equilibrium. The yellow circles (magenta squares) mark the locations 
on the curves where time $t$ = 1 second (1000 seconds). }
\label{fig:ntopratio} 
\end{figure}

We can understand the BBN yields shown in Figure \ref{fig:gravity} as
follows. For small values of $G$, the expansion rate is slow, and the
freezing of the weak interactions occurs later in cosmic history. As a
result, protons and neutrons remain in Nuclear Statistical Equilibrium
(NSE) longer and the $n/p$ ratio is smaller. As $G$ increases, the
expansion rate increases, freeze-out of weak interactions occurs
earlier, and the $n/p$ ratio is larger.  Figure \ref{fig:ntopratio}
shows the ratio $n/p$ as a function of temperature for different
values of the gravitational constant. The curves for different values
of $G$ show the basic trend outlined above, where weaker gravity
allows the $n/p$ ratio to track its NSE value (shown as the black
dashed curve) to lower temperatures, thereby resulting in lower
neutron abundances during the BBN epoch. At sufficiently late times,
free neutrons decay, and $n/p\to0$.  Note that the temperature
decreases with increasing time, but the relation $T_{\rm cm}(t)$
depends on the value of the gravitational constant. This trend is
illustrated in Figure \ref{fig:ntopratio}, where the circles mark the
location on the curves where time $t=1$ sec, and the squares delimit
$t=1000$ sec.  Since most of the neutrons are processed into helium-4,
its abundance generally grows with increasing $G$. The abundances of
deuterium and helium-3 also increase. As noted above, for sufficiently
large values of $G$, the abundance of helium-4 decreases again. A 
partial explanation is provided by Figure \ref{fig:ntopratio}, which 
shows that the $n/p$ ratio is not monotonic with increasing strength 
of gravity. The value of $n/p$ at the end of the BBN epoch (right 
side of the figure) increases as $G/G_0$ increases from unity to 100, 
but then decreases with further increase in $G/G_0$ from 100 to $10^6$. 
This decrease in the number of neutrons leads to less helium production 
for large values of $G/G_0$. 

\begin{figure}
\begin{center}
\includegraphics[scale=0.70]{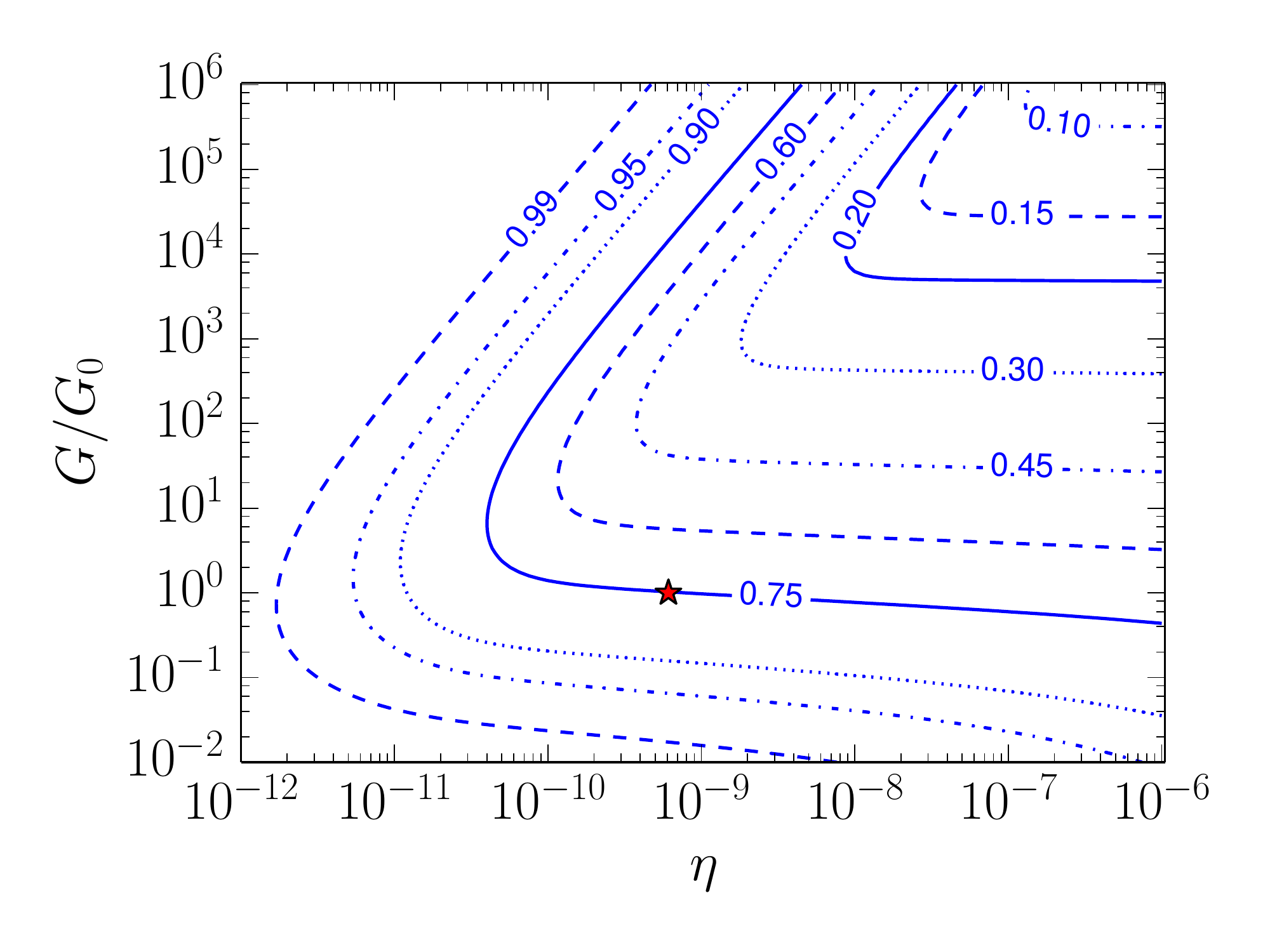} 
\end{center}
\vskip-1.0truecm 
\caption{Contours of the remaining Hydrogen mass fraction (including 
all isotopes) after BBN for universes with varying $\eta$ and $G/G_0$.
The thick contour marked 0.75 corresponds to hydrogen abundances
comparable to that of our universe. The red star marks the location of
our universe in the plane. The hydrogen abundance falls below 10
percent for the upper right part of the diagram, where $G/G_0\sim10^6$
and $\eta\sim10^{-6}$. }
\label{fig:bbnplane} 
\end{figure}

\subsection{Generalized Constraints from BBN} 
\label{sec:morebbn} 

Figure \ref{fig:bbnplane} shows the abundance of hydrogen in the dual
parameter space of $G/G_0$ versus $\eta$.  We plot contours of
constant mass fraction for all hydrogen isotopes, $X_{\rm H}$,
i.e., the summation of the single-proton hydrogen and deuterium mass
fractions 
\be
  X_{\rm H} = X_{^1{\rm H}} +X_{^2{\rm H}}.
\ee 
Figures \ref{fig:density} and \ref{fig:gravity} both show that when we
maximize either $\eta$ or $G/G_0$ while holding the other parameter
fixed, hydrogen results as the most abundant element.  We would expect
that BBN would produce a majority of hydrogen when we maximize both
parameters.  Figure \ref{fig:bbnplane} shows the opposite behavior:
$X_{\rm H}$ falls to below $10\%$ in the upper right-hand corner.  The
error in the logic resides in the phasing of the freeze-out of two
sets of reactions: the weak interactions responsible for setting the
ratio of neutrons to protons (denoted $n/p$); and the strong and
electromagnetic nuclear reactions responsible for the synthesis of
elements with mass number $A>1$.  There are six weak interactions
which dictate $n/p$, schematically shown as the forward and reverse
reactions
\begin{align}
  \nu_e+n&\leftrightarrow p+e^-,\label{eq:nue_on_n}\\
  e^+ +n&\leftrightarrow p+\overline{\nu}_e,\label{eq:pos_on_n}\\
  n&\leftrightarrow p+e^-+\overline{\nu}_e,\label{eq:ndecay}
\end{align}
where $e^\pm$ denotes positrons and electrons, $\nu_e$ denotes the 
electron neutrino, and $\overline{\nu}_e$ denotes the electron
antineutrino. As long as the rates associated with the reactions in
equations\ (\ref{eq:nue_on_n}) -- (\ref{eq:ndecay}) are rapid, the
ratio $n/p$ will stay in equilibrium and decrease as the universe
expands and the temperature decreases.  The lepton-capture rates
corresponding to the forward and reverse reactions in
equations\ (\ref{eq:nue_on_n}) and (\ref{eq:pos_on_n}) both scale as
the fifth power of temperature \cite{grohsfuller}.  The Hubble
expansion rate scales as the second power of temperature, so that 
the lepton-capture rates will always be larger than the expansion
rate at early times, and become subsidiary at later times.  If we
increase the strength of the gravitational constant, we precipitate an
earlier epoch when the expansion rate surpasses the lepton capture
rates.  The result is that $n/p$ goes out of equilibrium earlier,
dictating a larger free neutron abundance at this time. 

Figure \ref{fig:gravity} shows that the end state of most free
neutrons is their provisional incorporation into $^4{\rm He}$ nuclei.
For $1\lesssim G/G_0\lesssim100$, the rate of decrease in 
$X_{^1{\rm H}}$ is close to the rate of increase in $X_{^4{\rm He}}$.
However, Figure \ref{fig:gravity} also shows a local minimum for
$X_{^1{\rm H}}$ when $G/G_0\simeq100$.  At this point, the Hubble
expansion rate is fast enough that the reactions for synthesis of
$^4{\rm He}$, primarily the strong reaction 
$^3{\rm He}(^3{\rm He},2p)^4{\rm He}$, begin to freeze-out at an
earlier epoch.  The overabundance of neutrons either resides in other
nuclei, specifically deuterium (D) and $^3{\rm He}$, or continues to
exist as free neutrons.  The forward rate in equation
\eqref{eq:ndecay} is free neutron decay which is invariant with
temperature at late times.  Any remaining free neutrons at the
conclusion of BBN will transmute to free protons (see Figure
\ref{fig:ntopratio}).  Figure \ref{fig:gravity} shows that $^1{\rm H}$
mass fraction increases at a larger absolute rate than those of D and
$^3{\rm He}$, implying that $n/p$ decreases.  The mass fractions of D
and $^3{\rm He}$ are continuing to increase at the end of the
parameter range in Figure \ref{fig:gravity}, but they will reach a
local maximum eventually, much like the extant maxima for $^4{\rm He}$
and $^7{\rm Li}$.  To preserve the overabundance of neutrons which
exist after the freeze-out of the lepton capture rates, the strong and
electromagnetic rates must be more rapid.  Those rates are
proportional to varying powers of $\eta$, depending on the number of
input particles in any particular reaction.  Conversely, the rates for
the reactions in equations\ (\ref{eq:nue_on_n}) -- (\ref{eq:pos_on_n})
are insensitive to $\eta$, implying that $n/p$ will be roughly the
same for all $\eta$ at the epoch when the lepton-capture rates freeze
out.  The nuclear reactions are more rapid with increasing $\eta$,
implying a later epoch of nuclear freeze-out, implying a longer span
of time for the assembly of the larger nuclei.  Figure
\ref{fig:bbnplane} shows for any value of $G/G_0$, increasing $\eta$
increases $n/p$, or conversely decreases $X_{\rm H}$.

In any case, the results of Figure \ref{fig:bbnplane} show that the
mass fraction of hydrogen remaining after the epoch of BBN drops to
below 0.10 in the upper corner of the diagram, where $\eta\sim10^{-6}$
and $G/G_0\sim10^6$. Although some protons must remain after BBN in
order for the universe to have the raw material for water, the minimum
mass fraction is not known. Here we use 10 percent as a benchmark
value.  The resulting upper limit for the gravitational constant,
$G/G_0<10^6$ is comparable to that obtained by requiring stars to
function \cite{adams,adamsnew}. The value of $\eta$ can be larger than
that in our universe by a factor of more than 1000.

Although we use the 10 percent hydrogen abundance as a benchmark for
habitable universes, it is of interest to consider the location of the
contour for $X_{\rm H}=0$ in an extended version of the parameter
space shown in Figure \ref{fig:bbnplane}. To answer this question, we
consider the neutron to proton ratio $n/p$ in chemical equilibrium, 
\be
n/p = \exp\left(-\frac{\delta m_{np}}{T} + 
\frac{\mu_e}{T} - \xi_{\nu_e}\right),
\ee 
where $\delta m_{np}\sim1.3\,{\rm MeV}$ is the mass difference between
a neutron and a proton, $\mu_e$ is the chemical potential of the
electrons, and $\xi_{\nu_e}=\mu_{\nu_e}/T_{\rm cm}$ is the electron
neutrino degeneracy parameter \cite{grohsetal}.  The parameter 
$\xi_{\nu_e}$ is a co-moving invariant if we scale the electron
neutrino chemical potential with a temperature-like quantity different
from the plasma temperature.  In this setting, $T_{\rm cm}$ is the
co-moving temperature parameter and allows for the neutrinos to be out
of thermal equilibrium with the primeval plasma \cite{grohsetal}.  
Both $\mu_e/T$ and $\xi_{\nu_e}$ are small compared to $\delta
m_{np}/T$, so $n/p$ is strictly less than unity.  Figures
\ref{fig:density} and \ref{fig:gravity} show that for $n/p<1$, the
proton excess is mainly preserved in $^1{\rm H}$.  Therefore, under
the assumptions that $\mu_e/T$ and $\xi_{\nu_e}$ are small, the
$X_{\rm H}=0$ contour is never reached.  The contours in Figure
\ref{fig:bbnplane} continue to be spaced at larger intervals when both
$\eta$ and $G/G_0$ are increased.  The contour for $X_{\rm H}=0$ would
therefore be an asymptote assuming $\mu_e/T$ and $\xi_{\nu_e}$ are
small.  

The assumptions outlined above may not hold for every possible
universe within the multiverse. The electron chemical potential is
proportional to the baryon asymmetry, characterized by $\eta$.  If
$\eta$ sufficiently large, of order $\delta m_{np}/T\sim1$, then the
ratio $n/p$ could be larger than unity. However, for $\eta$ close to
unity, BBN would occur under matter-dominated conditions (see the
following subsection) where inhomogeneities are important
\cite{applegate,jedamzik}. This scenario is much different than the
parameter space considered in Figure \ref{fig:bbnplane}, where the
background medium is homogeneous and radiation-dominated.
Alternatively, if the parameter $\xi_{\nu_e}$ is large and negative,
i.e., if there exists an overabundance of antineutrinos to neutrinos,
then BBN continues to occur in radiation-dominated conditions with
$n/p>1$.  In this case we would expect very little $^1{\rm H}$ and
possibly a significant fraction of nuclei heavier than $^4{\rm He}$
--- depending on the specific value of $n/p$. These alternate
scenarios should be explored in future work, but are beyond the scope
of this present paper.

\subsection{Ordering of Time Scales and the Ratio of Densities}
\label{sec:ordering} 

In order for the universe to produce structure within the paradigm of
the standard cosmological model \cite{kolbturn}, the epoch of equality 
of matter and radiation must occur after the epoch of big bang
nucleosynthesis. Since we are not considering variations in the 
nuclear properties, the energy of nuclear reactions must be of order 
a few MeV. We can parameterize this level by writing 
\be
T_{\rm bbn} = b m_e\,,
\ee
where $m_e$ is the electron mass (0.511 MeV) and $b$ is a 
constant of order unity. The ordering constraint implies that 
\be
\tempeq < T_{\rm bbn} \qquad {\rm or} \qquad 
\eta {\Omega_M \over \Omega_b} < b {m_e \over \mpro} 
\sim 10^{-3} \,. 
\label{tconstraint} 
\ee
As discussed above, if we require the universe to retain at least 10
percent of its hydrogen after BBN, then the baryon to photon ratio
$\eta<10^{-6}$ (see Figure \ref{fig:bbnplane}).  With the value for
$\eta$ and the ratio $\Omega_M/\Omega_b\approx6$ found in our
universe, the epoch of BBN falls well before that of equality.  

In the bound on the energy density of the vacuum, given by equation
(\ref{vacbound}), the product $\eta(\Omega_M/\Omega_b)$ appears on the
right-hand-side.  If we set $\eta=10^{-6}$ in order to evaluate the
bound on $\rhovac$, then equation (\ref{tconstraint}) implies that the
largest value of the density ratio must be $\Omega_M/\Omega_b\approx100$, 
which is larger than the value in our universe by a factor of $\sim15$.

The bound of equation (\ref{conzero}) also indicates that the time of
vacuum domination $\timevac$ must occur after the time of matter
domination $\timeeq$. This result, in conjunction with that of equation
(\ref{tconstraint}), indicates that the time scales must obey the
ordering
\be
\timebbn < \timeeq < \timevac \,. 
\ee

\section{Conclusion} 
\label{sec:conclude}  

The main result of this paper is an upper bound on the density
$\rhovac$ of the vacuum energy, given by equation (\ref{vacbound}),
which represents a generalization of previous treatments.  If we
evaluate the right-hand-side of the equation using the values for
$(Q,\eta,\mplanck)$ found in our universe, the result is roughly
comparable to the ratio $\rhovac/\mplanck^4$ that is observed. This
finding is essentially a restatement of the coincidence problem -- in
our universe structure has formed recently and the energy density is
now becoming dominated by the vacuum. This paper shows (Section
\ref{sec:evaluate}) that a viable alternate universe could have larger
fluctuation amplitude $Q$ (by a factor of $\sim10^3$), larger baryon
to photon ratio $\eta$ (by a factor of $\sim10^3$), and stronger
gravity (so that $\mplanck$ is smaller by a factor of $\sim10^3$).
With the maximum value of $\eta$, the ratio of densities
$\Omega_M/\Omega_b$ can be larger by a factor of $\sim15$. With these
generalizations, the density of the vacuum energy can be much larger
than that of our universe: The constraint takes the form
$\rhovac/\mplanck^4<10^{-90}$ so that $\rhovac$ could be larger than
the observed value by $\sim30$ orders of magnitude.

This generalization of the bound on $\rhovac$ is significant. As
emphasized in the original paper of Weinberg \cite{weinberg87}, if the
maximum allowed value for the energy density $\rhovac$ is much larger
than the empirically allowed value, ``then we would have to conclude
that the anthropic principle does not explain why the cosmological
constant is as small as it is''. Given that $\rhovac$ could be
$\sim30$ orders of magnitude larger, relative to the Planck scale, and
still allow structure to form, we argue that the anthropic principle
has limited predictive power in this context.

A related outcome of this work is the finding that universes have
difficulty processing all of their protons into heavier elements, even
under extreme conditions. More specifically, we have considered
variations in both $\eta$ and $G$, and determined the fraction of
nucleons that are synthesized into non-hydrogen isotopes (Figure
\ref{fig:bbnplane}). The hydrogen mass fraction remains greater than 10
percent even when the parameters $\eta$ and $G$ are larger than $10^3$
and $10^6$ times the values realized in our universe. Moreover, the
results from Section \ref{sec:morebbn} indicate that some hydrogen
almost always remains after the epoch of BBN. In order to process all
of the protons into heavier elements, the neutron to proton ratio must
exceed unity ($n/p>1$), which in turn requires extreme conditions
(e.g., matter dominated BBN, large $\nu_e$ degeneracy parameter,
and/or large $\eta\sim1$). These results show that the BBN epoch does
not represent an example of fine-tuning. The input parameters can be
varied by many orders of magnitude and still allow the universe to
remain viable.

The results of this paper have implications for the broader issue of
the possible fine-tuning of the universe. The main result of this
paper is that the density of the vacuum energy $\rhovac$ could be
larger by more than 30 orders of magnitude and still allow for
structure formation (with appropriate adjustments to other
parameters). Of course, the density $\rhovac$ could also be 30 orders
of magnitude smaller and the universe would continue to develop cosmic
structure. We thus argue that the value of $\rhovac$ need not be
overly fine-tuned for the universe to produce observers.

The results of this paper also show that the epoch of Big Bang
Nucleosynthesis generally does not leave the universe with no
hydrogen. Specifically, the mass fraction of hydrogen remains larger
than 10 percent even when the baryon to photon ratio $\eta$ is larger
by a factor of $10^3$ and the gravitational constant $G$ is larger by
a factor of $10^6$.  The density ratio $\Omega_M/\Omega_b$ can be
larger by a factor of $\sim15$ if $\eta$ takes on its maximum value,
but can be much larger for smaller values of $\eta$. As a result, the
universe can remain chemically viable without fine-tuning the
parameters of the BBN epoch.

Finally, we have used previous results to constrain the other relevant
parameters of the problem. The amplitude $Q$ of the primordial density
fluctuations spans a range of a factor $\sim10^4$, from 10 times
smaller \cite{tegrees,tegmark} to $10^3$ times larger \cite{coppess}.
Additional considerations of stellar structure show that working stars
can exist for values of $G$ and the fine-structure constant $\alpha$
that vary over many orders of magnitude \cite{adams}. This claim holds
up in the face of additional constraints, including that the stars
have sufficiently hot surface temperatures and long lifetimes
\cite{adamsnew}, and that stars can form within their parental
galaxies \cite{reesost,whiterees}. Taken together, all of these
results indicate that the universe is not overly fine-tuned, in that
both the constants of physics $(\alpha,G)$ and the cosmological
parameters $(Q,\eta,\rhovac,\Omega_M/\Omega_b)$ can vary over a wide
range and still allow the universe to develop astrophysical structures
and perhaps even life.

\acknowledgments

We would like to thank I. Dell'Antonio, A. Erickcek, 
and J. Levicky for useful discussions. This work was supported
by JT Foundation through Grant ID55112: 
{\sl Astrophysical Structures in other Universes}, 
and by the Bahnson Trust Fund.


\end{document}